\documentclass{article}

\usepackage{amsfonts} \usepackage{amsmath}

\begin{document}

\title{New Isotropic and Anisotropic Sudden Singularities}

\author{John D. Barrow\thanks{e-mail address:
j.d.barrow@damtp.cam.ac.uk} and Christos G. Tsagas\thanks{e-mail
address: c.tsagas@damtp.cam.ac.uk}\\ {\small DAMTP, Centre for
Mathematical Sciences, University of Cambridge, Wilberforce
Road}\\ {\small Cambridge~CB3~0WA, UK}}

\date{\empty}

\maketitle

\begin{abstract}
We show the existence of an infinite family of finite-time
singularities in isotropically expanding universes which obey the
weak, strong, and dominant energy conditions. We show what new
type of energy condition is needed to exclude them ab initio. We
also determine the conditions under which finite-time future
singularities can arise in a wide class of anisotropic
cosmological models. New types of finite-time singularity are
possible which are characterised by divergences in the time-rate
of change of the anisotropic-pressure tensor. We investigate the
conditions for the formation of finite-time singularities in a
Bianchi type $VII_{0}$ universe with anisotropic pressures and
construct specific examples of anisotropic sudden singularities in
these universes.
\end{abstract}

\section{Introduction}
%%%%%%%%%%%%%%%%%%%%%%
It has recently been shown~\cite{jb,jb4} that the Einstein
equations permit the development of finite-time singularities
during the evolution of an expanding universe that obeys the
strong energy condition. In isotropic and homogeneous cosmologies
this singularity is characterised by a divergence in the pressure,
$p$, and the acceleration of the expansion scale factor $\ddot{a}$
as $t\rightarrow t_{s}<\infty$. The density $\rho$, expansion
scale factor $a,$ and expansion rate $\dot{a}/a$ all remain finite
at $t_{s}$. In order to exclude this type of singularity it is
sufficient to introduce a pressure-boundedness condition, for
example that $p<C\rho$ for some finite positive constant $C$, or
to require that ${\rm d}p/{\rm d}\rho$ be
continuous~\cite{jb,jb2}. Without such a condition linking the
pressure to the density, a pressure singularity can occur
independently of the density and leads to a violation of the
dominant energy condition (expressed by $|p|\leq\rho$) as
$t\rightarrow t_{s}$~\cite{jb,lake}. Note that the divergence of
$p$ with finite $\rho $ leads to an unbounded quantity with
dimensions of a velocity. Similar forms of singularity can be
obtained in Friedmann universes arising as cosmological solutions
of higher-order gravity theories~\cite{jb3,jb4} and other forms of
finite-time singularity have been proposed in the presence of
cosmological accelerations of fluid flow relative to the
hypersurface-orthogonal fluid congruence in~\cite{cot}. Sudden
singularities are sp curvature singularities that are neither
strong-curvature nor crushing~\cite{jb2} and their modest effects
on geodesics have been studied in~\cite{geo}. These mathematical
features of simple cosmological models impinge upon investigations
of the late-time behaviour of the so called 'phantom matter'
models. The latter generally have unconventional behaviour leading
to future 'big rip' singularities at finite time more severe than
those discussed here~\cite{rip}. It should be noted, however, that
not all phantom cosmologies end in a big rip
singularity~\cite{McI}.

We recall that, in the case of isotropic universes, sudden
singular behaviour results from a scale factor of the
form~\cite{jb}
\begin{equation}
a(t)=\left(\frac{t}{t_s}\right)^{q}\left(a_s-1\right)+ 1
-\left(1-\frac{t}{t_s}\right)^{n}\,,  \label{sol2}
\end{equation}
with $a_s\equiv a(t_s)$. Hence, as $t\rightarrow t_s$ from below,
we have
\begin{equation}
\ddot{a}\rightarrow q(q-1)Bt^{q-2}-
\frac{n(n-1)}{t_s^{2}[1-(t/t_s)]^{2-n}}\rightarrow-\infty\,,
\label{Lim}
\end{equation}
whenever $1<n<2$ and $0<q<1$. This solution of the Friedmann
equations exists in the interval $0<t<t_{s}$. More generally, a
finite-time singularity will arise in a Friedmann universe
when~\cite{jb4}
\begin{equation}
a(t)=\left(\frac{t}{t_s}\right)^{q}\left(a_s-1\right)+ 1-
(t_{s}-t)^{n} \left\{\sum_{j=0}^{\infty}\sum_{k=0}^{N_{j}}
a_{jk}(t_s-t)^{j/Q}(\log^{k}[t_s-t])\right\}\,,  \label{gen}
\end{equation}
where $a_{jk}$ are constants, $N_j\leq j$ are positive integers,
$Q\in\mathbb{Q}^{+}$ and the quantity in \{..\} brackets is a
convergent double psi series. The latter tends to zero as
$t\rightarrow t_s$.

In this paper we extend previous work on the subject by studying a
new family of sudden singularities in isotropic Friedmann
universes and the formation of sudden singularities in the
presence of anisotropic expansion. In~\cite{dab} an anisotropic
and inhomogeneous cosmology of Stephani type was found to possess
similar finite-time singularities, as well as spatial analogues
that preserve the dominant energy condition within finite regions
of space. We shall investigate whether new forms of anisotropic
finite-time singularity can arise, possibly under weaker energy
conditions than in the isotropic case.

We begin our discussion with a covariant formulation of the sudden
singularity problem in section 2. This allows us to identify the
various possibilities that may occur.  In section 3 we describe a
new infinite family of sudden singularities in isotropic universes
which obey all the standard energy conditions. In section 4 we
give an explicit example of some of the possibilities uncovered in
section 2 by constructing anisotropic sudden singularities in a
Bianchi type $VII_0$ universe with anisotropic pressures. We
summarise with our conclusions and a discussion of finite-time
singularities in section 5.

\section{Covariant formulation}
%%%%%%%%%%%%%%%%%%%%%%%%%%%%%%%
We begin by investigating under what conditions new types of
finite-time singularity can arise in an anisotropic, irrotational,
spatially homogeneous, spatially-flat spacetimes, containing
matter with density $\rho$, isotropic pressure $p$ and a
trace-free anisotropic pressure tensor $\pi_{ab}$. The evolution
of this model is described by the following set of covariant
equations (e.g.~see~\cite{EvE})
\begin{eqnarray}
\dot{\rho}&=&-\Theta(\rho+p)- \sigma_{ab}\pi^{ab}\,, \label{edc}\\
\dot{\Theta}&=&-{\textstyle{1\over3}}\Theta^{2}-
{\textstyle{1\over2}}(\rho+3p)- 2\sigma^{2}\,,  \label{Ray}\\
\dot{\sigma}_{ab}&=&-{\textstyle{2\over3}}\Theta\sigma_{ab}-
\sigma_{c\langle a}\sigma^{c}{}_{b\rangle}- E_{ab}+
{\textstyle{1\over2}}\pi_{ab}\,,  \label{sigmadot}\\
\dot{E}_{ab}&=&-\Theta E_{ab}-
{\textstyle{1\over2}}(\rho+p)\sigma_{ab}-
{\textstyle{1\over2}}\dot{\pi}_{ab}-
{\textstyle{1\over6}}\Theta\pi_{ab}+ 3\sigma_{c\langle
a}E^{c}{}_{b\rangle} \notag\\ &{}&
-{\textstyle{1\over2}}\sigma_{c\langle a}\pi^{c}{}_{b\rangle}\,.
\label{Edot}
\end{eqnarray}
Here, $\Theta$ is the volume expansion scalar, $\sigma_{ab}$ is
the shear tensor, describing kinematical anisotropies, and
$E_{ab}$ is the electric part of the Weyl tensor, which is
associated with tidal forces. We proceed by assuming that $p$ is
independent of $\rho$ and that $\pi_{ab}$ is independent of both
$p$ and $\rho$. We will also leave the time evolution of
$\pi_{ab}$ unspecified by any constitutive relationships. These
propagation equations must be supplemented by the constraint
equations
\begin{eqnarray}
\Theta^{2}&=&3(\rho+\sigma^{2})\,,  \label{Fried}\\
\dot{\sigma}_{ab}&=&-\Theta\sigma_{ab}+ \pi_{ab}\,,  \label{GC1}
\end{eqnarray}
both imposed by the spatial flatness of the model. Note that the
first of these is the generalised Friedmann equation and the
latter is the residual Gauss-Codacci formula, which here provides
an alternative description of the shear evolution. Also,
expressions (\ref{sigmadot}) and (\ref{GC1}) combine to give
\begin{equation}
\pi_{ab}= {\textstyle{2\over3}}\Theta\sigma_{ab}-
2\sigma_{c\langle a}\sigma^{c}{}_{b\rangle}- 2E_{ab}\,,
\label{GC2}
\end{equation}
which directly relates all the sources of anisotropy.

Introducing non-zero spatial curvature leaves
Eqs.~(\ref{edc})-(\ref{sigmadot}) the same, although generally it
adds a $\mathrm{curl}H_{ab}$ term to the right-hand side of
(\ref{Edot}) (where $H_{ab}$ is the magnetic Weyl tensor). In the
presence of curvature Eqs.~(\ref{Fried}) and (\ref{GC1}) also
change into
\begin{equation}
\mathcal{R}=2\left(\rho-{\textstyle{1\over3}}\Theta^{2}+
\sigma^{2}\right)\,,  \label{cFried}
\end{equation}
and
\begin{equation}
\mathcal{R}_{\langle ab\rangle}=-\dot{\sigma}_{ab}-
\Theta\sigma_{ab}+ \pi_{ab}\,,  \label{cGC1}
\end{equation}
respectively. Here, $\mathcal{R}_{\langle ab\rangle}$ is the
symmetric and trace-free component of $\mathcal{R}_{ab}$, the
spatial Ricci tensor, $\mathcal{R}=\mathcal{R}^a{}_a$ is the
associated Ricci scalar and $\sigma^{2}=\sigma_{ab}\sigma^{ab}/2$
is the magnitude of the shear tensor. Also, Eq.~(\ref{cGC1})
combines with (\ref{sigmadot}) to give
\begin{equation}
\mathcal{R}_{\langle ab\rangle}=
-{\textstyle{1\over3}}\Theta\sigma_{ab}+ \sigma_{c\langle
a}\sigma^c{}_{b\rangle}+ E_{ab}+ {\textstyle{1\over2}}\pi_{ab}\,.
\label{cGC2}
\end{equation}

According to the system (\ref{edc})-(\ref{GC2}), we can have an
\textit{anisotropic-pressure singularity} where the quantities
$a$, $\Theta$, $\rho$, $p$ and $\sigma_{ab}$ are all finite but
$\pi_{ab}$ diverges as $t\rightarrow t_{s}<\infty$. Here the
singular behaviour is manifested as a finite time singularity in
$\dot{\sigma}_{ab}$, the first time derivative of the shear tensor
(see Eqs.~(\ref{sigmadot}), (\ref{GC1}) and
(\ref{cGC1})).\footnote{One may also consider the vacuum case by
setting $\rho=0=p=\pi_{ab}$. Then, the reduced system of
Eqs.~(\ref{edc})-(\ref{cGC2}) appears to allow for a finite time
singularity in $\dot{\sigma}_{ab}$ if ${\cal R}_{\langle
ab\rangle}$ or $E_{ab}$ diverge. However, when $\Theta$ is bounded
within the time interval in question, a series of known existence
theorems due to Wald and Rendall prevents this from
happening~\cite{W}.} Let us now look at this possibility from a
different perspective.

Consider the scale factor $a$, defined by means of the isotropic
volume expansion, according to $\dot{a}/a=\Theta/3$. Then,
Raychaudhuri's formula (see Eq.~(\ref{Ray})) transforms into
\begin{equation}
\frac{\ddot{a}}{a}=-{\textstyle{1\over6}}(\rho+3p)-
{\textstyle{2\over3}}\sigma^{2}\,.  \label{Ray1}
\end{equation}
Applied to a Bianchi~$I$ spacetime, the time derivative of the
above gives
\begin{equation}
\frac{\dddot{a}}{a}=-{\textstyle{1\over9}}\Theta\rho-
{\textstyle{1\over2}}\dot{p}+
{\textstyle{5\over9}}\Theta\sigma_{ab}\sigma^{ab}-
{\textstyle{1\over2}}\sigma_{ab}\pi^{ab}\,,  \label{dddota}
\end{equation}
given the spatial flatness of the model. In an anisotropic
cosmology with non-zero spatial curvature, however, this
generalises to
\begin{equation}
\frac{\dddot{a}}{a}=-{\textstyle{1\over9}}\Theta\rho-
{\textstyle{1\over2}}\dot{p}+
{\textstyle{5\over9}}\Theta\sigma_{ab}\sigma^{ab}-
{\textstyle{1\over2}}\sigma_{ab}\pi^{ab}+
{\textstyle{2\over3}}\sigma_{ab}\mathcal{R}^{\langle ab\rangle}\,,
\label{cdddota}
\end{equation}
where $\mathcal{R}_{\langle ab\rangle}$ satisfies expression
(\ref{cGC2}). Following (\ref{Ray1})-(\ref{cdddota}), we see that
$\dddot{a}$ diverges when $\pi_{ab}$ is not bounded. At the same
time the lower-order derivatives of the scale factor remain
finite. It worth noticing that, unless appropriate regularity
conditions are imposed on the matter, a diverging $\dot{p}$ will
also lead to a singularity in $\dddot{a}$ (see
Eq.~(\ref{cdddota})).

When dealing with anisotropic expansion it helps to consider the
scale factor defined by means of the generalised Hubble law. In
covariant terms the latter reads~\cite{E}
\begin{equation}
\frac{\dot{\ell}}{\ell}=\frac{\dot{a}}{a}+
\sigma_{ab}e^{a}e^{b}\,,  \label{dotell}
\end{equation}
where $e_{a}$ are unitary, linearly independent, spacelike vectors
(i.e.~$e_{a}e^{a}=1$ and $e_{a}u^{a}=0$). The shear term in the
right-hand side of the above ensures that, in contrast with $a$,
the new scale factor also contains information on the anisotropy
of the expansion. The full role of the anisotropy is revealed by
taking the time derivative of (\ref{dotell}). On using
(\ref{sigmadot}) the latter leads to
\begin{equation}
\frac{\ddot{\ell}}{\ell}=\frac{\ddot{a}}{a}+
\left(\sigma_{ab}e^{a}e^{b}\right)^2- \left(\sigma_{c\langle
a}\sigma^c{}_{b\rangle}+E_{ab}-{\textstyle{1\over2}}\pi_{ab}\right)
e^{a}e^{b}\,,  \label{ddotell}
\end{equation}
with $\ddot{a}/a$ satisfying expression (\ref{Ray1}). In addition,
one can use expression (\ref{dotell}) to define the individual
scale factors along the three linearly-independent directions
determined by $e_{i}$ (with $i=1,2,3$). For example, contracting
$\sigma_{ab}$ twice along $e_{1}$ in \ref{dotell}) we obtain
\begin{equation}
\frac{\dot{\ell_{1}}}{\ell_{1}}=\frac{\dot{a}}{a}+ \sigma_{11}\,,
\label{dotelli}
\end{equation}
which defines the scale factor in the $e_{1}$-direction.
Similarly, Eq.~(\ref{ddotell}) leads to
\begin{equation}
\frac{\ddot{\ell}_{1}}{\ell_{1}}= \frac{\ddot{a}}{a}+
\sigma_{11}^{2}- \sigma_{c\langle 1}\sigma^{c}{}_{1\rangle}-
E_{11}+ {\textstyle{1\over2}}\pi_{11}\,,  \label{ddotelli}
\end{equation}
with analogous expressions for $\ell_{2}$ and $\ell_{3}$. Results
(\ref{ddotell}) and (\ref{ddotelli}) show that, when $\pi_{ab}$,
$E_{ab}$ or $\mathcal{R}_{\langle ab\rangle}$ diverges,
$\ddot{\ell}$ and $\ddot{\ell}_{i}$ will generally diverge as
well, which is in distinct contrast with the behaviour of $a$.
Recall that an unbounded anisotropic pressure leads to the
singular behaviour of $\dddot{a}$ instead of $\ddot{a}$ (see
Eqs.~(\ref{Ray1}), (\ref{dddota})). Note also that, unlike a
diverging $\pi_{ab}$, a diverging $\dot{p}$ manifests itself as a
singularity in the third time derivative of $\ell$ and not in the
second.

So far, our inspection of the self-consistent divergences in the
sets (\ref{edc})-(\ref{GC2}), (\ref{Ray1})-(\ref{cdddota}) and
(\ref{dotell})-(\ref{ddotelli}) has provided necessary rather than
sufficient conditions for these new types of finite-time
singularities to occur. Clearly, their detailed form must also
comply with all of the Einstein field equations. In what follows
we will construct explicit examples to show how some of the
aforementioned situations may arise.

\section{New isotropic examples}
%%%%%%%%%%%%%%%%%%%%%%%%%%%%%%%%
It is clear from Eq.~(\ref{dddota}) that in the isotropic case
($\sigma_{ab}=0$) there exists an infinite family of Friedmann
sudden singularities with the scale factor $a(t)$ chosen to have
the form (\ref{sol2}), or its generalisation (\ref{gen}), which
satisfy the weak energy condition ($\rho\geq0$ and $\rho+p\geq0$),
the strong energy condition ($\rho+3p\geq0$ and $\rho+p\geq0$),
and the dominant energy condition ($\rho\geq0$ and $\rho\pm
p\geq0$). If we choose $n$ in Eq.~(\ref{sol2}) so that
\begin{equation*}
N<n<N+1\,,
\end{equation*}
where $N$ is a positive integer, then there will be a finite-time
singularity in the $(N+1)^{st}$ time derivative of the scale
factor. In other words,
\begin{equation*}
\frac{{\rm d}^{N+1}a}{{\rm d}t^{N+1}}\rightarrow\infty
\hspace{5mm} \text{as} \hspace{3mm} t\rightarrow t_s\,,
\end{equation*}
with all the lower order derivatives remaining finite, namely
\begin{equation*}
\frac{{\rm d}^ra}{{\rm d}t^{r+1}}\rightarrow L<\infty \hspace{5mm}
\text{as} \hspace{3mm} t\rightarrow t_{s}\,,
\end{equation*}
for all $r=1,2,\ldots\,N$. In this limit the $(N+1)^{st}$ time
derivative of the scale factor produces an infinity in the
$(N-1)^{st}$ time derivative of the pressure at $t_s$, but the
density $\rho$ and all lower derivatives of $p$ remain finite as
$t\rightarrow t_s$ since
\begin{equation*}
\frac{1}{a}\;\frac{{\rm d}^{N+1}a}{{\rm d}t^{N+1}}\rightarrow-
{\textstyle{1\over2}}\;\frac{{\rm d}^{N-1}p}{{\rm
d}t^{N-1}}\rightarrow- \infty \hspace{5mm} \text{as} \hspace{3mm}
t\rightarrow t_s\,.
\end{equation*}
Specifically, for a scale factor of the form (\ref{sol2}), we see
that
\begin{equation*}
\frac{{\rm d}^{N+1}a}{{\rm d}t^{N+1}}\rightarrow
(-1)^{N}\frac{n(n-1)(n-2)\cdots(n-N)}
{t_s^{N}[1-(t/t_s)]^{N+1-n}}+\,\, {\cal O}
\left(\frac{t}{t_s}\right)^{q-1-N}\,.
\end{equation*}

If we choose $N=1$, we create the sudden singularities given
in~\cite{jb,jb4} and also by Eq.~(\ref{Lim}) above. Because they
arise in $\ddot{a}$ and $p$ they lead to a violation of the
dominant energy condition despite satisfying the weak and strong
energy conditions. However, the infinite family of sudden
singularities characterised by the choice $N\geq2$ satisfy the
weak, strong and dominant energy conditions and produce no
divergent spacelike energy fluxes of the type discussed
in~\cite{lake}. Unbounded behaviour occurs in the time derivatives
of the pressure and might suggest the need to define a family of
new energy conditions if this behaviour is to be excluded by fiat.
For example, we could exclude finite-time singularities of this
type if we introduced a \textit{generalised matter regularity
condition that required}
\begin{equation*}
\frac{{\rm d}^rp}{{\rm d}t^r}<C_s\frac{{\rm d}^s\rho}{{\rm
d}t^s}\,, \hspace{5mm} \text{for some} \hspace{2mm} s\leq r\,,
\end{equation*}
where $C_s$ is a positive constant.

\section{Anisotropic examples}
%%%%%%%%%%%%%%%%%%%%%%%%%%%%%%
Let us look in more detail at some of the possibilities revealed
in the last section. Consider the spatially homogeneous,
anisotropically expanding, Bianchi~$VII_0$ universe with a line
element given by~\cite{L1,L2,BS}
\begin{equation}
\mathrm{d}s^2=-\mathrm{d}t^2+
g_{\alpha\beta}\mathrm{d}x^{\alpha}\mathrm{d}x^{\beta}\,,
\label{VII01}
\end{equation}
where $t$ is the comoving proper time and
\begin{equation}
g_{\alpha \beta }=\left(
\begin{array}{c}
A^2[\cosh\mu+\sinh\mu\cos(kz)] \hspace{10mm}
A^2\sinh\mu\sin(kz) \hspace{10mm} 0 \\
A^2\sinh\mu\sin(kz) \hspace{10mm}
A^2[\cosh\mu-\sinh\mu\cos(kz)] \hspace{10mm} 0 \\
\hspace{15mm} 0 \hspace{34mm} 0 \hspace{34mm} B^2
\end{array}
\right)  \label{VII02}
\end{equation}
with $\alpha$, $\beta=1,2,3$ and $z\equiv x^3$; also, $A=A(t)$,
$B=B(t)$, $\mu=\mu(t)$ and $k$ is a constant parameter. The above
metric can be interpreted as the superposition of an axisymmetric
Bianchi~I spacetime and a circularly polarised gravitational wave
propagating along the axis of symmetry (see~\cite{L1,L2} and
also~\cite{BS}). The wavenumber and amplitude of this wave is
given by $k$ and $\mu$ respectively. Note that the above metric
has anisotropic spatial curvature as well as anisotropic expansion
and reduces to the axisymmetric Bianchi~$I$ spacetime in the
long-wavelength limit (i.e.~as $k\rightarrow0$). It is the most
general spatially homogeneous generalisation (or perturbation) of
the zero-curvature Friedmann universe and has played an important
role in discussions of the isotropy of the universe~\cite{CH}.

When matter is in the form of a fluid with anisotropic pressures,
the Einstein equations associated with the metric (\ref{VII01}),
(\ref{VII02}) lead to the evolution formulae~\cite{L2,BS}:
\begin{equation}
\frac{\ddot{A}}{A}+\frac{\dot{A}}{A}
\left(\frac{\dot{A}}{A}+\frac{\dot{B}}{B}\right)=
{\textstyle{1\over2}}(\rho-p_3)\,,  \label{Addot}
\end{equation}
\begin{equation}
\frac{\ddot{B}}{B}+2\,\frac{\dot{A}}{A}\frac{\dot{B}}{B}=
{\textstyle{1\over2}}(\rho+p_3-p_1-p_2)+
{\textstyle{1\over2}}\left(\frac{k}{B}\right)^2\sinh^2\mu\,,
\label{Bddot}
\end{equation}
and
\begin{equation}
\ddot{\mu}+\left[2\left(\frac{\dot{A}}{A}\right)
+\frac{\dot{B}}{B}\right]\dot{\mu}+
{\textstyle{1\over2}}\left(\frac{k}{B}\right)^2\sinh(2\mu)=
p_1-p_2\,.  \label{muddot}
\end{equation}
These are supplemented by the following constraint:
\begin{equation}
\left(\frac{\dot{A}}{A}\right)\left[\frac{\dot{A}}{A}
+2\left(\frac{\dot{B}}{B}\right)\right]
-{\textstyle{1\over4}}\left[\dot{\mu}^2
+\left(\frac{k}{B}\right)^2\sinh^{2}\mu\right]=\rho\,.
\label{Fried1}
\end{equation}
which is the generalised Friedmann equation. The above reduces to
the standard Friedmann equation for the flat FRW metric, where
$A=B$ and $\mu=0$. In Eq.~(\ref{Fried1}) $A=A(t)$ and $B=B(t)$
represent the two scale factors that characterise the anisotropic
expansion. Also recall that $\mu=\mu (t)$ and $k$ are respectively
the amplitude and the wavenumber of the superimposed gravitational
wave of wavelength $2\pi B/k$.

Consider now the following power-law evolution equations for the
two individual scale factors
\begin{eqnarray}
A(t)&=&A_s- 1+ \left(\frac{t}{t_s}\right)^{q}-
\mathcal{A}(t_s-t)^{n}\,,  \label{A}\\
B(t)&=&B_s- 1+ \left(\frac{t}{t_s}\right)^{m}+
\mathcal{B}(t_s-t)^r\,,  \label{B}
\end{eqnarray}
where $t_s$ corresponds to a time in the late evolution of the
model, $A_{s}=A(t_s)$, $B_s=B(t_s)$ and $\mathcal{A}$,
$\mathcal{B}$ are constants. Also, $q$, $m\in\mathbb{\mathbb{R}}$
and, for reasons that will become clear next, we will assume that
$2<n$ and that $1<r<2$. Then, taking the first and second time
derivatives of $A$ we find
\begin{eqnarray}
\dot{A}&=&\frac{q}{t_s}\left(\frac{t}{t_s}\right)^{q-1}+
n\mathcal{A}(t_s-t)^{n-1}\,,  \label{dotA}\\
\ddot{A}&=&\frac{q(q-1)}{t_s^2}\left(\frac{t}{t_s}\right)^{q-2}-
n(n-1)\mathcal{A}(t_s-t)^{n-2}\,,  \label{ddotA}
\end{eqnarray}
while by successively differentiating $B$ we obtain
\begin{eqnarray}
\dot{B}&=&\frac{m}{t_s}\left( \frac{t}{t_s}\right)^{m-1}-
r\mathcal{B}(t_s-t)^{r-1}\,,  \label{dotB} \\
\ddot{B}&=&\frac{m(m-1)}{t_s^2}\left(\frac{t}{t_s}\right)^{m-2}+
r(r-1)\mathcal{B}(t_s-t)^{r-2}\,.  \label{ddotB}
\end{eqnarray}
Given that $2<n$, Eqs.~(\ref{dotA}) and (\ref{ddotA}) guarantee
that all of $A$, $\dot{A}$, and $\ddot{A}$ remain finite as $t$
approaches $t_{s}$. On the other hand, since $1<r<2$, expressions
(\ref{dotB}), (\ref{ddotB}) imply that $\ddot{B}$ diverges as
$t\rightarrow t_s$, while $B$ and $\dot{B}$ remain finite. In
addition, since $\ell=(AB^2)^{1/3}$ is the average scale factor of
the anisotropic expansion, both $\ell $ and $\dot{\ell}$ are
finite at the $t\rightarrow t_s$ limit but
$\ddot{\ell}\rightarrow\infty$. All these properties ensure that
at $t=t_s$ the model described by Eqs.~(\ref{A}) and (\ref{B})
experiences a sudden singularity at $t_s$ of a sort discussed in
section 2. Assuming the metric (\ref{VII01}), (\ref{VII02}), this
sudden singularity is triggered by a divergent anisotropic
pressure at $t_s$. For example, when one or both of the principal
pressures $p_1$ and $p_2$ diverge as $t\rightarrow t_s$,
expression (\ref{Bddot}) guarantees that $\ddot{B}$ also diverges
at $t_s$. Note that the singular behaviour of $\ddot{B}$ is in
agreement with the power-law evolution of (\ref{B}). At the same
time, Eq.~(\ref{Addot}) guarantees that $\ddot{A}$ remains finite,
which is compatible with definition (\ref{A}). Also, according to
(\ref{muddot}), the divergence of either $p_1$ or $p_2$
immediately implies a singular behaviour for $\ddot{\mu}$, unless
$p_{1}=p_{2}$. Clearly, the rest of the variables, namely $\rho$,
$p_3$, $\mu$ and $\dot{\mu}$, do not show any singular behaviour
and all the terms in Eq.~(\ref{Fried1}) are always finite at
$t>0$. Since $\rho$ remains finite at $t_s$, while at least one of
the principal pressures diverges there, leading also to a
divergence of $\dot{\rho}$, we conclude that in our example the
dominant energy condition is violated at the $t=t_s$
limit~\cite{jb,lake}.

Alternatively, we may assume that both $p_1$ and $p_2$ are finite,
and allow for $p_3$ to diverge as $t\rightarrow t_s$. Then,
Eqs.~(\ref{Addot}) and (\ref{Bddot}) ensure that both $\ddot{A}$
and $\ddot{B}$ diverge at $t_s$, while (\ref{muddot}) shows that
$\ddot{\mu}$ remains bounded at that point. This type of singular
behaviour is also described by the scale factors (\ref{A}) and
(\ref{B}), with the extra proviso that now $n$ must also lie
within the open interval $(1,2)$.

It is straightforward to generalise these examples in the same way
that we did with isotropic singularities in the last section. By
choosing $n$ and $r$ to lie in appropriate open intervals bounded
by successive positive integers we can produce mutatis mutandis an
infinite family of sudden singularities that occur in arbitrarily
high derivatives of one or both of the anisotropic scale factors
and in the associated time derivatives of $\pi_{ab}$.

Finally, we note that the anisotropic scale factors given in
Eqs.~(\ref{A})-(\ref{B}) can be generalised in the same way that
the particular isotropic solution (\ref{sol2}) was generalised
using the logarithmic series to expression (\ref{gen})
in~\cite{jb4}. More specifically, we have
\begin{eqnarray}
A(t)&=&A_s- 1+ \left(\frac{t}{t_s}\right)^{q}-
\mathcal{A}(t_s-t)^n\left\{\sum_{j=0}^{\infty}\sum_{k=0}^{N_j}
a_{jk}(t_s-t)^{j/Q}(\log^k[t_s-t])\right\}\,, \notag\\
B(t)&=&B_s- 1+ \left(\frac{t}{t_{s}}\right)^{m}+
\mathcal{B}(t_s-t)^r\left\{\sum_{j=0}^{\infty}\sum_{k=0}^{M_j}
b_{jk}(t_s-t)^{j/S}(\log^k[t_s-t])\right\}\,, \notag
\end{eqnarray}
where $m,n,q,$ and $r$ are as before and $a_{jk}$, $b_{jk}$ are
constants, $N_j,M_j\leq j$ are positive integers and
$Q,S\in\mathbb{Q}^+$.

\section{Conclusions}
%%%%%%%%%%%%%%%%%%%%%
We have shown how a new infinite family of finite-time
singularities can arise in isotropically expanding universes which
obey the strong, weak and dominant energy conditions. This is
possible because they produce infinities in the time derivatives
of the pressure of arbitrarily high order. In particular,
singularities in ${\rm d}^{N+1}a/{\rm d}t^{N+1}$ arise from
singularities in ${\rm d}^{N-1}p/{\rm d}t^{N-1}$ for all positive
integers, $N$. These can only be excluded by introducing new
regularity conditions on the derivatives of the pressure.

We have also shown that finite-time singularities can arise in new
ways when the expansion of the universe is anisotropic. The simple
forms of sudden singular behaviour found in isotropic universes
are not rendered unstable by expansion anisotropies. Moreover, we
have found that new types of finite-time singularity can arise
from the anisotropic part ($\pi_{ab}$) of the fluid pressure. We
investigated the role of anisotropic pressure in detail. Beginning
with a general discussion of the issue of anisotropic finite-time
singularities, we identified a number of possible singular cases.
We constructed a specific anisotropic model of Bianchi type
$VII_{0}$ to show how an unbounded anisotropic pressure tensor can
lead to finite-time singularities. The latter can occur whilst the
matter density and isotropic pressure remain finite and can be
avoided only by introducing appropriate restrictions on
$\pi_{ab}$. For example, one might require that
$\pi_{ab}\pi^{ab}<C\rho^2$ for some positive constant $C$. In
summary, our investigations have shown that finite-time
singularities are a feature of a wide class of cosmological
solutions to the Einstein equations and are not simply a
consequence of special symmetries. In order to exclude them from
studies of the general behaviour of the Einstein equations at late
times we need to introduce new restrictions on the allowed form of
the energy-momentum tensor.

\section*{Acknowledgments}
%%%%%%%%%%%%%%%%%%%%%%%%%%
We would like to thank Brett McInnes,  Shin'ichi Nojiri and
Hongwei Yu for helpful comments and Alan Rendall for drawing our
attention to refs.~\cite{W}.

\end{document}